# Phase partitioning in a novel near equi-atomic AlCuFeMn alloy

Dr. Amritendu Roy, Dr. Mainak Ghosh, Dr. Partha Sarathi De[1]

**Abstract**

A novel low cost, near equi-atomic alloy comprising of Al, Cu, Fe and Mn is synthesized using arc-melting technique. The cast alloy possesses a dendritic microstructure where the dendrites consist of disordered FCC and ordered FCC phases. The inter-dendritic region is comprised of ordered FCC phase and spinodally decomposed BCC phases. A Cu segregation is observed in the inter-dendritic region while dendritic region is rich in Fe. The bulk hardness of the alloy is ~ 380 HV, indicating significant yield strength.

Traditionally, metallic alloys are designed to have one principal element with multiple other elements in small quantity. This design philosophy is predicated on the premise that such compositional strategy promotes a predominantly single phase ductile matrix where strengthening is achieved by a fine distribution of $2^{nd}$ phase or solute atoms. High entropy (HE) alloys [1-2] have been suggested as a conceptually new class of material where unlike conventional alloys, multiple elements (usually 4-5 or more) are added in equimolar or near-equimolar concentrations. Contrary to the expectation of multiple phases, HE alloys have a predominantly single phase body centered cubic (BCC) or face centered cubic (FCC) structure and fascinating structural and functional properties [3,4]. Most of these HE alloys however comprises of costly transition metals like Co, Ni, Nb, Mo etc. limiting their commercial viability [5]. The present work, proposes a simple four component equiatomic alloy system comprising of Al, Cu, Fe and Mn, in a near equiatomic ratio.

---

[1] Communicating author

To predict the phase formation in HE alloys, a common design principle in use is the parameter based constraints of difference in atomic size ($\delta$) and enthalpy of mixing ($\Delta H_{mix}$) [6],

$$\delta = \sqrt{\left[\sum_{i=1}^{N} c_i \left(1 - \frac{r_i}{\sum_{i=1}^{N} c_i r_i}\right)^2\right]} \quad (1)$$

$$\Delta H_{mix} = \sum_{i=1, i \neq j}^{N} 4 \Delta H_{AB}^{mix} c_i c_j \quad (2)$$

where $c_i$ and $r_i$ represents the molar content and atomic radius of the i[th] component, $\Delta H_{AB}^{mix}$ the mixing enthalpy for the binary system A- B and R being the universal gas constant. Small atomic size differences ($\delta < 4\%$) and mixing enthalpy (-10 kJ/mole $<\Delta H_{mix}<5$ kJ/mole) combined with high mixing entropy ($\Delta S_{mix} > 13.38$ JK$^{-1}$ mol$^{-1}$) is suggested to stabilize the single phase solid solution [6]. Guo et al. [7] proposed an additional Valence Electron Concentration (VEC) parameter,

$$VEC = \sum_{i=1}^{n} c_i . n_i \quad (3)$$

where $n_i$ is the average number of itinerant electrons per atom. A VEC value $\geq 8$ is suggested to favor a FCC structure and the BCC structure stable at lower values [7]. For Al, Cu, Fe, Mn the atomic radii, and average itinerant electrons per atom are given in table-1 and mixing enthalpies for the binary combinations in table 2. For the proposed alloy this predicts an atomic size difference of ~6%, the enthalpy of mixing as -7 kJ/mole and a VEC of 7.25. Consequently a mixture of FCC and BCC phase is expected in the given alloy. The phases present in the suggested alloy is further experimentally verified and compared with the theoretical prediction. The transformations occurring in this novel alloy and its possible effect on mechanical properties are also discussed.

**Table 1:** The atomic radii and average itinerant electrons for elements present in AlCuFeMn [3].

| Elements | Al | Cu | Fe | Mn |
|---|---|---|---|---|
| Atomic radii | 1.432 | 1.28 | 1.241 | 1.350 |
| Itinerant electrons | 3 | 11 | 8 | 7 |

**Table 2:** The binary enthalpy of mixing for elements present in AlCuFeMn [8].

| $\Delta H_{mix}$ | Al | Cu | Fe | Mn |
|---|---|---|---|---|
| Al |  | -1 | -11 | -19 |
| Cu |  |  | 13 | 4 |
| Fe |  |  |  | 0 |

High purity Al, Cu, Fe and Mn (all purity > 99.95%) in equimolar proportions is arc melted (Ti gettered to remove oxygen) to produce an alloy button of ~ 5 gm weight. The sample button is homogenized by melting and re-melting it 6 times, flipping it at each stage. Sample from the homogenized specimen is metallographically polished and subsequently etched with Glyceregia for further investigation using Optical Microscopy and Field Emission Scanning Electron Microscopy (FESEM). The FESEM imaging is done in both topographical mode and in compositional mode at an acceleration voltage of 5 kV. The differential interference contrast image of etched microstructure (Fig. 1a) reveal islet like areas i.e. Region A distributed within a continuous matrix network i.e. Region B, typical of cellular microstructure where Region A are the cells and Region B is the inter-dendritic region. At high magnifications the inter-dendritic Region B is observed to possess a distinct periodic threadlike interlaced structure while Region A has a globular structure (Fig. 1b). Elemental mapping using Energy Dispersive Spectroscopy (EDS) technique indicate a partitioning of Fe into Region A and a Cu rich Region B (Figure 2 a to c)). The Region A constitutes ~64.4 ± 6 % volume fraction while Region B comprises of ~35.6 ± 6 % of the microstructure.

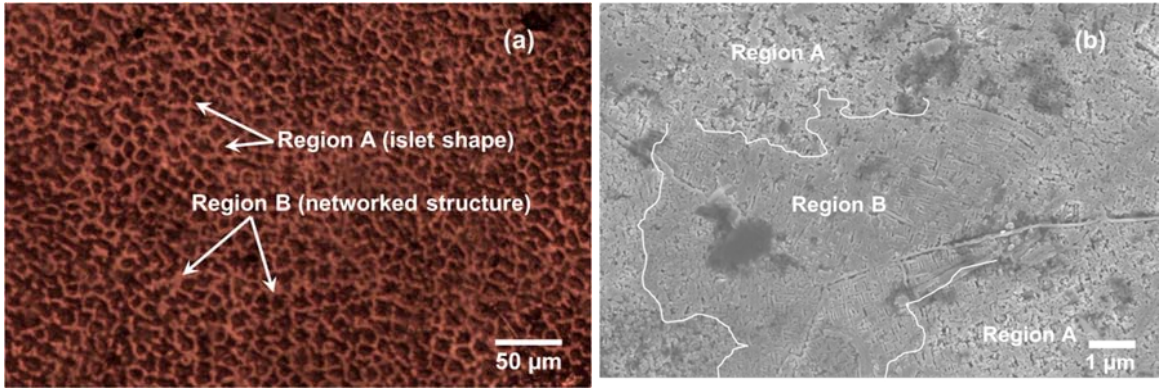

**Fig. 1.** (a) Optical image of the etched HE alloy in (b) SEM image of the dendritic and inter-dendritic regions.

In table -3 the overall specimen composition as well as average composition of dendritic and inter-dendritic Region-B is provided. The composition of Region-A obtained using spot EDS is more or less uniform at all locations of the specimen. However, spot EDS results of Region-B show compositional variation, with Cu content reaching ~ 40% at some locations, with a commensurate decrease in Fe content. A smaller fluctuation in Mn content is also observed.

To further confirm the phase identity at Region A and B, transmission electron microscopy examination of the specimen is done. Perforated thin foils are prepared by twin jet electro-polisher using a mixture of ethanol: perchloric acid at sub-ambient temperature. The prepared specimens are examined in analytical transmission electron microscope.

**Table 3:** Composition of the Regions A and B region measured using spot EDS along with overall composition measured by area EDS in FESEM..

| Phase | At % Cu | At % Mn | At % Fe | At% Al |
| --- | --- | --- | --- | --- |
| Average Comp | 30.6 ± 4.2 | 22.2 ± 0.8 | 24.2 ± 0.9 | 23.0 ± 2.6 |
| Region A | 23.3 ± 1.4 | 23.9 ± 0.3 | 29.1 ± 0.1 | 23.7 ± 1.5 |
| Region B | 40.2 ± 9.7 | 19.7 ± 3.4 | 17.1 ± 4.8 | 23.0 ± 1.5 |

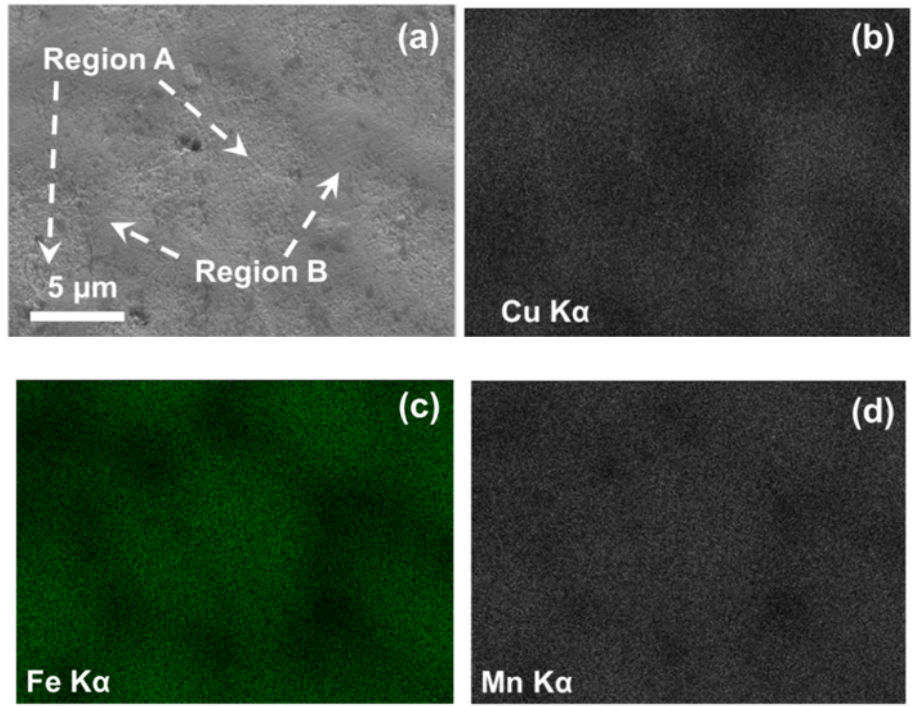

**Fig. 2.** (a) The phase distribution under SEM and corresponding (b)EDS map of Cu (c) EDS map of Fe and (d) EDS map of Mn. The EDS map of Al (not shown here) is uniform throughout.

Based on the morphology, TEM investigations detected three types of structures (a) the small nearly globular (~2-15 nm) structures within dendritic Region A (b) rounded edge irregular / oval shaped (~20-50 nm) geometry within dendritic Region A and (c) elongated threaded rounded end with high aspect ratio in inter-dendritic Region B. The small globular structures shown by arrow (Fig. 3a) exhibited nearly same contrast under different beam direction indicating nearly homogeneous structure and is identified as FCC disordered structure with lattice parameter, $a$ ~2.57 Å. These small precipitates are invisible in the SEM image (Fig. 1b) unlike the relatively larger oval phase in Region A which also possesses a FCC structure (Fig. 3c). The Selected Area Diffraction (SAD) pattern of this oval phase (Fig. 3d) indicated the presence of extra row of spots (shown by white arrows) nearly at the mid-way of principal rows, i.e. super lattice reflection signifying a ordered FCC structure with lattice parameter, $a$ ~5.73 Å (Fig. 3d). Another feature observed is the presence of numerous parallel lines with spacing of 0.5-2.0 nm within the oval shaped structures in Region A (Fig. 3c). These are identified as nano-twins, the presence of which is confirmed from the analyzed diffraction pattern (Fig. 3d). Arc melting in copper mold is expected to result in a non-equilibrium cooling rate creating residual stresses within the button specimen. The nano-twins formed are possibly an upshot of the stresses present in the specimen.

The threaded inter-dendritic structure (Fig. 3e) i.e. Region-B is observed to possess a disordered BCC crystallography and a lattice parameter of $a$~3.01 Å. The selected area diffraction pattern from Region-B (Fig. 3f) endorsed not only the signature of BCC phase, but also of ordered second phases (lattice parameter a ~ 5.73 Å) in between resulting in the appearance of secondary spots. The region in between the oval shaped FCC phase maintains an orientation relation of $[001]_{FCC}$ ∥ $[011]_{BCC}$ with the BCC. The results indicate that the inter-dendritic Region B is probably a eutectic region where the liquid has solidified into the pro-eutectic FCC phase (ordered) and additional phases. The threaded net like structure observed is typical of spinodal decomposition reaction where composition fluctuation proceeded in an elastically soft direction [9]. For cubic crystals this is generally in the <100> direction [9]. In fact, similar threaded structures observed by Tong et al. in $Al_xCoCrCuFeNi$ alloys are ascribed to spinodal decomposition [10]. The work of Santodonato et al. on same HE alloy system indicates occurrence of spinodal decomposition into a BCC/B2 combination with additional FCC phase in between [11]. In this work, the crystal structure of the other spinodally decomposed phase (i.e. in between the threads) is not separately identifiable. However, high resolution EDS performed on TEM specimens show a Cu segregation into the threaded region (Fig. 4a). This indicates that the crystal structure and lattice parameter of phase between the threaded zone is in all probability similar to the disordered BCC phase except for a difference in chemical composition. The volume fraction of inter-thread region is however much lower than the disordered BCC phase being ~ 10% or lesser compared to the threaded zone.

To identify the detailed crystal structure of the major phase's present a room temperature powder X-ray diffraction (pXRD) using CuKα over a 2θ range of 20° - 140° is done. This data is used for further structural refinement studies using the constant scale factor (Le-Bail) and Rietveld Refinement technique where an open source software FullProf [12] is used. Different possible crystal structure combinations evaluated using Le-Bail refinement indicates a combination of FCC, $Fm\bar{3}m$ (SG # 225) and BCC $Im\bar{3}m$ (SG # 229) as the best fit result. The slight difference in the lattice parameter measured by XRD with respect to the TEM data for BCC and FCC phase is due to the inherent inaccuracies in TEM measurement. The best fit results of the Le-Bail refinement is used as an initial input for the Rietveld Refinement approach which along with the average composition obtained from EDS studies on TEM sample (Table 3) is used to determine the site occupancy of the atoms. The atomic positions within the unit cell of two phases including their relative phase fractions as evaluated by Rietveld refinement is given in Table 4. The goodness of fit of the Rietveld study with the experimental XRD results is shown in Fig. 4(c). The refinement results predict that the BCC and FCC phase are approximately in 1:2 proportions and agrees well with the volume fraction measured for Region A and B using image analysis. While the occupation by the alloying elements at different lattice sites in the BCC phase is unambiguous (Single inequivalent site (Wyckoff position, 2a), occupied by the constituting elements), the situation is complex for the FCC phase. In this phase, constituent alloying elements can occupy one or more of 4a, 4b and 8c sites with partial or complete site occupation indicating ordering of the FCC phase which confirms the TEM results. The presence of 3 different phases, indicates that the prediction of a BCC and FCC phase using δ, $\Delta H_{mix}$ and $VEC$ parameters is not applicable in the present work.

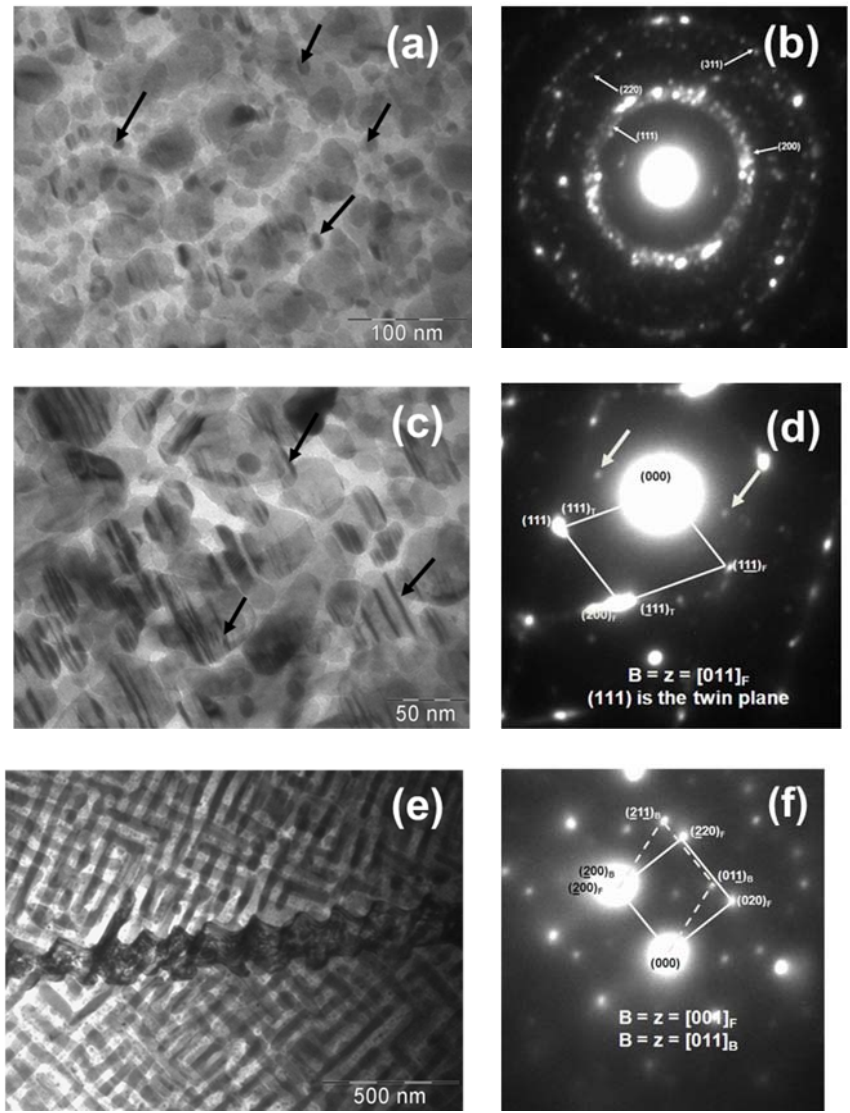

**Fig. 3.** (**a**) Bright field image showing distribution of near globular fine scale second phase (**b**) Polycrystalline ring pattern associated with the globular phase in "a", (**c**) irregular / elliptical shaped ordered second phase containing twins (**d**) SAD pattern associated with (c), (**e**) Bright field image showing threaded structure of Region B (see Fig. 1a) (**f**) SAD pattern of (e).

A preliminary assessment of the microstructure for mechanical property indicates that the finely distributed ordered FCC phase is expected to produce order and solute strengthening. This along with the spinodally decomposed region will result in significant grain boundary strengthening. As the sample volume in this work is too small for a bulk mechanical tensile test an estimate of the yield strength is obtained from the bulk hardness obtained as 380 ±3 HV. Thus, the AlCuFeMn alloy developed in this work is expected to have yield strength $(\sigma_{y.s.})$ of ~ 950 MPa assuming that $\sigma_{y.s.}$ is 2.5 times that of hardness value [13].

**Table 4**: Refined structural parameters of BCC (Im-3m) and FCC (Fm-3m) obtained from Rietveld refinement of the powder X-ray diffraction of AlCuFeMn multicomponent alloy. Conventional Rietveld parameters are: $R_p$ = 26.3; $R_{wp}$ = 19.8 and $R_{exp}$ = 10.3.

| Phase: $Im\bar{3}m$ (SG#229) <br> Lattice Parameters: a = b = c = 2.9285 Å <br> α = β = γ = 90º <br> Relative Phase Fraction: 36.72 % <br> Composition: Cu = 39.00 at.%; Fe = 18.00 at.%; Al = 23.00 at.% and Mn = 20.00 at.% | | | | | | Phase: $Fm\bar{3}m$ (SG#225) <br> Lattice Parameters: a = b = c = 5.8589 Å <br> α = β = γ = 90º <br> Relative Phase Fraction: 63.28 % <br> Composition: Cu = 23.75 at.%; Fe = 29.50 at.%; Al = 23.75 at.% and Mn = 23.00 at.% | | | | | |
|---|---|---|---|---|---|---|---|---|---|---|
| Overall alloy composition: Cu = 29.39 at.%; Fe = 25.25 at.%; Al = 23.47 at.% and Mn = 21.89 at.% | | | | | | | | | | |
| Atom | Wyckoff Pos. | x | Y | Z | Occ. | Atom | Wyckoff Pos. | x | y | z | Occ. |
| Cu | 2a | 0 | 0 | 0 | 0.39 | Cu <br> Fe | 4b | 0.5 | 0.5 | 0.5 | 0.95 <br> 0.05 |
| Fe | | | | | 0.18 | Fe | 4a | 0 | 0 | 0 | 1.00 |
| Al | | | | | 0.23 | Al | | | | | 0.475 |
| Mn | | | | | 0.20 | Mn <br> Fe | 8c | 0.25 | 0.25 | 0.25 | 0.460 <br> 0.065 |

In conclusion the following observations can be made,

(a) A novel low cost high entropy alloy comprising of AlCuFeMn is proposed. The alloy has a bulk hardness of ~ 380 HV, indicating significant yield strength.

(b) The AlCuFeMn alloy comprises of a combination of disordered BCC, ordered FCC and a disordered FCC phase.

(c) The microstructure reveals a dendritic structure where dendrites comprise of ordered/disordered FCC phase. The inter-dendritic region comprises of ordered FCC and BCC phases which assumes a spinodal dendrite form.

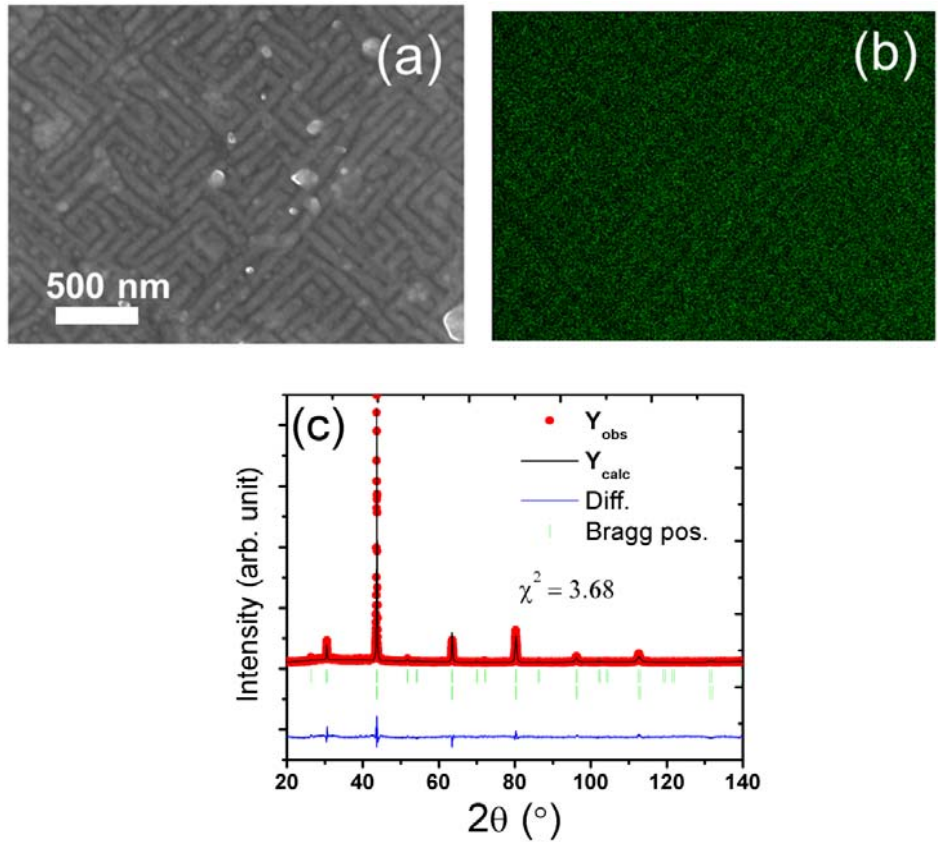

**Fig. 4.** (**a**) SEM image of the TEM specimen showing Region B (**b**) The Copper distribution map of Region B in EDS mapping showing copper concentration in the threaded BCC phase (**c**) Rietveld refinement result of AlCuFeMn sample (2θ = 20 to 140°).

**Acknowledgements:** One of the authors (PSD) acknowledges the help provided by Mr. SubhaSanket Panda and Mr. Ritukesh Sharma in sample polishing and etching.